\documentclass[print]{aastex}
\usepackage{graphicx}
\begin{document}
\title{Coronal density diagnostics with Si X: CHANDRA/LETGS observations of Procyon,
$\alpha$~Cen~A$\&$B, Capella and $\epsilon$~Eri}
\author{G. Y. Liang, G. Zhao, and J. R. Shi}
\affil{National Astronomical Observatories, Chinese Academy of
Sciences, \\ Beijing 100012, P. R. China} \email{gzhao@bao.ac.cn}


\begin{abstract}
Electron density diagnostics based on a line intensity ratio of Si
X are applied to the X-ray spectra of Procyon,
$\alpha$~Cen~A$\&$B, Capella and $\epsilon$~Eri measured with the
Low Energy Transmission Grating Spectrometer (LETGS) combined with
High-resolution Camera~(HRC) on board the {\it Chandra X-ray
Observatory}. The ratio $R_1$ of the intensities of the Si X lines
at 50.524~\AA\, and 50.691~\AA\, is adopted. A certain of the
temperature effect in $R_1$ appears near the low-density limit
region, which is due to the contamination of Si X line at
50.703~\AA\,. Using the emission measure distribution (EMD) model
derived by Audard et al. (2001) for Capella and emissivities
calculated with APEC model by Smith et al. (2001), we successfully
estimate contributions of Fe XVI lines at 50.367~\AA\, and
50.576~\AA\, (73\% and 62\%, respectively). A comparison between
observed ratios and theoretical predictions constrains the
electron densities (in logarithmic) for Procyon to be
8.61$^{+0.24}_{-0.20}$~cm$^{-3}$, while for $\alpha$~Cen~A$\&$B,
Capella and $\epsilon$~Eri to be 8.81$^{+0.27}_{-0.23}$~cm$^{-3}$,
8.60$^{+0.39}_{0.32}$~cm$^{-3}$, 9.30$_{-0.48}$~cm$^{-3}$ and
9.11$^{+1.40}_{-0.38}$~cm$^{-3}$, respectively. The comparison of
our results with those constrained by the triplet of He-like
carbon shows a good agreement. For normal stars, our results
display a narrow uncertainty, while for active stars, a relatively
larger uncertainty, due to the contamination from Fe XVI lines, is
found. Another possible reason may be that the determination of
the continuum level, since the emission lines of Si~X become weak
for the active stars. For $\epsilon$~Eri, an electron density in
the C V forming region was estimated firstly through Si X
emissions.
\end{abstract}


\section{Introduction}
Since the seminal work by Vaiana et al. (1981), it has become
clear that all late-type stars share the same basic coronal
characteristics: hot thermal plasma with temperatures around
1---10~MK covering stellar surfaces, magnetic confinement, the
presence of flares, etc. A systemic investigation indicates that
most active stars have X-ray luminosity up to 2 orders of
magnitude higher than that of solar corona (Scelsi et al. 2005;
Peres et al. 2004). Present popular assumption for interpretation
of such difference can be attributed to their different
composition in terms of various kinds of coronal structures
(ranging from the relatively faint and cool structures of the
background corona to the very bright and hot flaring regions) and
to the number of X-ray emitting coronal structures present.

The solar corona as observed with the modern X-ray and XUV
telescopes on board Yohkoh, SoHO, or TRACE, is found to be
extremely structured, and even in the high angular resolution
TRACE images, it appears to be spatially unresolved fine
structure. Yet spatially resolved X-ray observations of stellar
coronae are currently not feasible. Some information on the
spatial distribution of stellar coronae was inferred from X-ray
light curves of suitably chosen stars such as eclipsing binaries
(White et al. 1990; Schmitt $\&$ K${\rm \ddot{u}}$rster 1993;
G${\rm \ddot{u}}$del et al. 1995, 2003; Siarkowski et al. 1996),
but such analyses can only be carried out for very special systems
with advantageous geometries, and the actual information derivable
from such data is rather limited. Another method to infer the
structure in spatially unresolved data is spectroscopic
measurements of the electron density, that is, X-ray spectra allow
us to get the structure information for a various stellar coronae.
Nevertheless, the temperature distribution (or emission measure
distribution EMD) and the coronal abundance could be estimated
from low-resolution spectra by an application of global fitting
approaches. While the previous measurements did not allow
measuring density $n_e$ and emission measure {\it EM}
independently because the information from spectral lines was not
available from spectra with the low resolution, such that no
emitting volumes {\it V} could be estimated from {\it EM
=$n_e^2~V$}. Therefore the information about loop size wasn't
accessible.

Direct spectroscopic information on plasma densities at coronal
temperatures on stars other than the Sun firstly became possible
with the advent of ``high-resolution" spectra
($\lambda/\Delta\lambda \sim200$) obtained by the {\it Extreme
Ultraviolet Explorer} (EUVE) which is capable of separating
individual spectral lines. Even with this resolution, the
available diagnostics have often tended to be not definitive,
owing to the poor signal-to-noise ratio (SNR) of the observed
spectra or blended lines. After the launch of new generation
satellites {\it Chandra} and XMM-{\it Newton}, the high-resolution
spectra coupled with the large effective area has made it is
possible to measure individual lines in the X-ray range for a
large sample of stars in the same fashion as X-ray emission lines
from the solar corona obtained and analyzed for many years (Doyle
1980; McKenzie $\&$ Landecker 1982; Gabriel et al. 1988). The
emissivity of selected lines depends on the density. Some lines
may be present only in low-density plasmas, such as the forbidden
line in He-like triplet, while other lines may appear only in
high-density plasmas (such as lines formed following excitations
from excited levels).

Ness et al. (2002, 2004) systemically investigated the coronal
density using the characteristic of He-like triplet for stars with
various activity covering inactive and active levels. For the
hot-temperature region, the density is estimated from emission
lines of carbon-like iron, and a typical density ranging
10$^{11}$----10$^{12}$cm$^{-3}$ was obtained. For the
low-temperature region, the density can be derived from low-$Z$
elements with low ionization energies such as C V, which is lower
by at least an order of magnitude than that of the hot-temperature
region. A typical electron density ranging
10$^8$---10$^{10}$cm$^{-3}$ has been derived for solar and
solar-like coronae by authors (Audard et al. 2001, Brinkmann et
al. 2000). In inactive stars, emission lines of Si VII---Si XIII
have also been clearly detected, and some lines of Si X show a
high SNR in the wavelength range covered by LETGS. In the
collision ionization equilibrium (Mazzatto et al. 1998) condition,
the temperature of peak fractional abundance of Si X is very close
to that of C V. This means that they form in the same region and
share the same electron density. Therefore, the density derived
from Si X should be comparable to that from He-like C V. In our
recent work (Liang et al. 2005), we noticed that the line
intensity ratio $R_1$ of Si X at the lines at 50.524~\AA\, and
50.691~\AA\, originating from $3d\to2p$ transitions is sensitive
to the density, whereas it is insensitive to the temperature. So
an application of this ratio $R_1$ in several stellar coronal
spectra is performed.

In this paper, we derive the electron densities for stars:
Procyon, $\alpha$ Cen A$\&$B, Capella and $\epsilon$~Eri using
this ratio $R_1$ for the first time, and compare the derived
densities with those from He-like C V. The paper is structured as
follows: we present our sample and a detailed description of line
flux measurements in Sect. 2. A brief description of theory model
is introduced in Sect. 3. Diagnostic of the electron density and
discussions are presented in Sect. 4. The conclusions are given in
Sect. 5.

\section{Observations and data analyses}
The new generation of X-ray telescopes and X-ray spectrometers on
board {\it Chandra} and XMM-{\it Newton} has opened the world of
spectroscopy to the X-ray astronomy with high-resolution and high
effective collecting areas. Spectroscopic measurements can be
performed with instruments, such as the Reflection Grating
Spectrometer (RGS) on board XMM-{\it Newton}, and the High and
medium Energy Grating (HEG and MEG) spectrometers, and the Low
Energy Transmission Grating Spectrometer (LETGS) on board {\it
Chandra}. Though RGS, HEG, and MEG provide large effective areas
in the high energy region (5 to 40~\AA\,) with high-resolution,
the LETGS covers a much larger wavelength range 5---175~\AA\, with
the high spectral resolution. One specific advantage of the LETGS
resulting from its covered large range of wavelength is that the O
VII triplet at around 21~\AA\,, the C V triplet at around 41~\AA\,
and $3d\to2p$ transition lines at 50~\AA\, of Si X can be measured
in one spectrum. In this wavelength range, a number of emission
lines from highly ionized Si ions have been found in the coronal
spectra~(Raassen et al. 2002, 2003). Here, we pay a special
attention on the emission lines at 50.524~\AA\, and 50.691~\AA\,
of Si~X from $3d\to2p$ transitions.

Our sample consists of 5 stars including three normal dwarf stars,
i.e.,  Procyon and $\alpha$~Cen~A and B, an active late-type dwarf
star, $\epsilon$ Eri, and an active binary system Capella. The
properties of our sample, along with ObsID and exposure time are
summarized in Table 1. All observations adopt grating of LETGS
combined with HRC instrument on board {\it Chandra} Observatory.
In case of Capella, an additional observation (with ObsID=55) with
ACIS-S instrument is used. Because ACIS-S instrument has
significant energy resolution to separate overlapping spectral
orders, LETGS+ACIS-S observations are better choice for
determination of the electron density. However only one
observation is available from {\it Chandra Data Archival Center}
for our sample. Another goal of analysis for the ACIS-S spectrum
of Capella is to validate whether there is significant
contamination from high-order (refer to $m\geq2$) spectra around
the selected lines (50.524 and 50.691~\AA\,). If line fluxes
derived from the two different observations are comparable, no
contamination from high-order spectra can be concluded. This also
backs up our analyses for other stars. Reduction of the LETGS
datasets uses CIAO3.2 software with the science threads for
LETGS/HRC-S observations. For the ACIS-S spectrum of Capella, {\sf
Destreak} tool was used to clean the events induced by detector
artifacts on CCD 8 (``streaks"). For the extraction of the spectra
of $\alpha$~Cen A\&B, similar procedure as employed by Raassen et
al. (2003) is adopted here under the CIAO environment. Figure 1
shows the spectra with background subtracted for Procyon, $\alpha$
Cen A$\&$B, Capella and $\epsilon$~Eri in the wavelength range of
50---55~\AA\,.

\subsection{Determinations of line fluxes}
The line fluxes were derived by modelling the spectrum locally
with narrow Gaussian profiles, in which the instrumental effective
area extracted from CIAO3.2 calibration files with {\sf fullgarf}
tool has been folded, together with a constant representing the
background and (pseudo)continuum emissions which were determined
in the line-free region (51--52~\AA\,). Here, only the positive
order spectra has been adopted because there is a plate gap around
52~\AA\, for the negative spectra. The average instrumental FWHM
of lines is about 0.06~\AA\, for LETGS observations. The fluxes
have been obtained after correction for the effective area. In the
fitting, 1$\sigma$ uncertainty is adopted to determine statistical
errors for the line fluxes, and no wavelength correction has been
used to the dispersion relation of the HRC-S/LETG. The error bars
determined in this method are slightly underestimated due to the
adopted functional forms of the line profiles, which introduce a
certain of systematic errors. Table 2 shows the theoretical
wavelengths and measured fluxes of the two prominent lines of Si
X, together with results of Fe XVI lines detected in Capella and
$\epsilon$~Eri stars. Since the centered wavelength for a given
line is also a thawed parameter in the fitting, the best-fit
result is different for the different stars, yet it is consistent
with theoretical identification within 30m\AA\,. So only the
theoretical wavelength is given for each emission line in this
table.

\subsubsection{Procyon, $\alpha$ Cen A$\&$B}
For normal stars such as Procyon and $\alpha$ Cen A$\&$B, a
detailed description has been presented by Raassen et al. (2002,
2003) and a conclusion that the coronal plasmas are dominated by
plasmas with temperatures ranging 1---3~MK was obtained. $\alpha$
Cen A and B were firstly spectrally resolved from observation with
LETGS instrument, a detailed description about it is presented by
Raassen et al. (2003). In the long wavelength region $>30$~\AA\,,
$3d\to2p$ transitions of Si X at 50.524~\AA\, and 50.691~\AA\, are
the prominent lines, which ensures the correct measurements of
line fluxes. In order to measure the fluxes of the two lines
correctly, contamination from higher order spectral lines of high
ionized Fe ions should be taken into account. Fortunately,
emission from high charged Fe ions is either weak or absent in
first order (and particularly in third order which is damped by a
factor $\approx20$) for inactive stars. In addition, the
temperature of peak emissivity of Fe XVI lines ranging this
wavelength region is about 4~MK, which is higher than that of
X-ray emitting plasma. Such that no emission lines of Fe XVI are
detected for the three stars.

In order to obtain the line fluxes, two Gaussian profiles with the
same FWHM value, combining with a constant were used in fitting
the spectra in 50--51~\AA\,. The continuum level is around
1.2$\times10^{-4}$photon~s$^{-1}$cm$^{-2}$\AA\,$^{-1}$ for the
three normal stars, which is determined by fitting with a constant
model in line-free wavelength region 51--52\AA\,. The best-fit
values of line fluxes are listed in Table 2, along with statistic
errors (with percentage) derived with 1$\sigma$ uncertainty. The
observed wavelengths are within 10~m\AA\,, when compared with
experimental ones. The reduced $\chi^2$ are 0.76, 0.63 and 0.67,
respectively, which indicates present fitting is good for the
three normal stars.

\subsubsection{Capella}
For active star Capella, numerous studies exist in the literatures
for the active star Capella (Mewe et al. 2001, Canizares et al.
2000, Audard et al. 2001), which reported its properties using
observations with different instruments. Most observations adopted
this star as test platform to calibrate instruments on board
satellites (Brinkmann et al. 2000). The observation adopting the
ACIS-S instrument eliminates the contamination from high-order
spectral lines for the selected lines (Si X). In order to validate
whether there is contamination from the high-order spectral lines
for the selected Si~X lines, the observation with ACIS-S
instrument is also analyzed. Here, the -1 order spectrum is
adopted, because no photon has been detected in +1 order spectrum
above 30~\AA\,. Nevertheless, the contamination from lines of Fe
XVI around 50~\AA\, is inevitable, since they have become the
strongest lines as shown in Fig. 1. For much more active star, the
emissions of Si X completely disappear.

In the fitting, four components (with same FWHM) combined with
continuum emission
(5.7$\times10^{-4}$photon~cm$^{-1}$s$^{-1}$\AA$^{-1}$ determined
in the line-free region) were used for the ACIS-S and HRC spectra
in 50--51~\AA\,. The reduced $\chi^2$ for ACIS-S and HRC
observations are 0.75 and 0.71, respectively. Fig. 2 shows the
best-fit (dotted line) to the ACIS-S spectrum. In this plot, the
HRC spectrum of Capella is also overlapped (dashed histogram) for
the visual comparison though their effective areas are slightly
different. The best-fit results of the two different observations
indicate that there is no contamination from the high-order
spectra. The derived line fluxes are comparable each other, such
as for line at 50.367~\AA\,, the derived fluxes are 2.38$\pm$0.30
and 2.25$\pm$0.22$\times10^{-4}$phot.cm$^{-2}$s$^{-1}$ for ACIS
and HRC observations, respectively, for line at 50.691~\AA\,, they
are 0.98$\pm$0.24 and
1.14$\pm$0.18$\times10^{-4}$phot.cm$^{-2}$s$^{-1}$. Such
consistency gives us confidence to use the line fluxes derived
from HRC spectra for other stars, though no ACIS-S observation
data is available for our sample. In case of Capella, ACIS-S
spectrum was adopted for the analysis in the following.

In the spectra observed with ACIS-S and HRC, the line of Si X at
50.691~\AA\, has been clearly detected. Is the observed flux at
50.576~\AA\, completely from contribution of Si X at 50.524~\AA\,?
In the available database such as CHIANTI, MEKAL and APED, we note
that Fe XVI line at 50.555~\AA\, may also be the significant
source of contribution to the observed flux at 50.576~\AA\,,
because present instrument can not resolve the two emission lines
obviously. The most prominent lines at 50.367~\AA\,, 54.136~\AA\,
and 54.716~\AA\, in this range were identified to be emissions of
Fe XVI by Raassen et al. (2002), which further confirms that Fe
XVI contribute to the observed flux at 50.576~\AA\,. So an another
problem is that how much fraction of the flux at 50.576~\AA\,
comes from Si X emission at 50.524~\AA\,.

The emission measure distribution (EMD) of Capella has been
derived from different observations with different instruments
with high-resolutions such as HEG (Canizares et al. 2000), LETGS
(Mewe et al. 2001) on board {\it Chandra} and RGS (Audard et al.
2001) on board XMM-{\it Newton}. Their results indicate that the
EMD has a sharp peak around 7~MK, together with a smaller one
grouping around 1.8~MK which is necessary to explain the detected
O VII He-like triplet and C VI Ly$\alpha$ line. Argiroffi et al.
(2003) investigated its variability during different phase, and
found that the emission is compatible with a constant source. In
this work, we adopt the EMD derived from RGS observation by Audard
et al. (2001), with combination of emissivity of Fe XVI included
in (Astrophysical Plasma Emission Code) APEC model to derive the
contributions of Fe XVI lines. We found that Fe XVI line at
50.555~\AA\, contributes $\sim$62\% to the total observed
fluxes~(2.38$\pm$0.30$\times10^4$phot.cm$^{-2}$s$^{-1}$). The
remaining flux at this wavelength should come from Si X line at
50.524~\AA\,. Additionally, we notice that there is a shoulder
around 50.576~\AA\,. So we fit the spectrum again by adding an
additional component. We initially set two artificial components
around 50.57~\AA\,. The best-fit (smooth solid line in Fig. 2) of
the spectrum shows a better result than the fitting with four
components (dotted line) by the visual inspection. The reduced
$\chi^2$ is also more close to unity which changes from 0.75 to
0.79. Though the obtained flux
(2.07$\pm$0.28$\times10^4$phot.cm$^{-2}$s$^{-1}$) of Fe XVI line
at 50.576~\AA\, is slightly higher than the theoretical prediction
(1.65), the two values from different procedure are comparable.
The line flux of Si X line at 50.524~\AA\, listed in Table 2, is
from the fitting procedure.

For the line flux at 50.367~\AA\, we further note that about
$\sim$73\% contribution comes from Fe XVI line at 50.350~\AA\,. A
search of APEC shows that the remaining flux may be from
contribution of Fe XVII lines (50.342~\AA\, and 50.361~\AA\,) and
Si X (50.333~\AA\,, 50.336~\AA\, and 50.359~\AA\,). As the
strongest lines of Si X at 50.524~\AA\, appear to be weak in
Capella, we suggest that the remaining flux of the observed flux
at 50.367~\AA\, is from emission lines of Fe XVII. Using the EMD
model derived by Audard et al. (2001) and the emissivity of Fe
XVII predicted by the APEC model, we predict the total fluxes of
the four lines are about
0.68$\times10^{-4}$photon~cm$^{-2}$~s$^{-1}$, which is consistent
with the remaining flux
(0.97$\times10^{-4}$photon~cm$^{-2}$~s$^{-1}$) within statistical
error.

\subsubsection{$\epsilon$~Eri}
The detailed analysis for Capella reveals that the contamination
due to high-order spectral lines is negligible. However, the
emission from Fe XVI lines must be taken into account. Four
components and a constant value are used to fit the spectrum of
$\epsilon$~Eri in wavelength range 50--51~\AA\,. The continuum
emission determined in the line-free region is
1.2$\times10^{-4}$photon~cm$^{-1}$s$^{-1}$\AA$^{-1}$. The best-fit
(reduced $\chi^2$=0.62) results of line fluxes are listed in Table
2, along with 1$\sigma$ statistical error.

For the flux extraction of Si X line at 50.524~\AA\,, the similar
procedure as for Capella is applied to estimate the contribution
of Fe XVI lines. Using the emissivity of Fe XVI extracted from
APEC and normalization according to the isolated line at
54.134~\AA\,, we estimate that the contributions of Fe XVI to the
observed fluxes around 50.350~\AA\, and 50.555~\AA\, are about
53\% and 39\%, respectively.  For line flux around 50.550~\AA\,,
the remaining flux is from contribution of Si X line
(50.524~\AA\,).

\section{Theory model of line ratio}
In our previous paper (Liang et al. 2005), we describe properties
and application of Si X spectrum in detail, and found that six
line ratios are sensitive to electron density. One of these ratios
$R_1$ has a good application on density diagnostic because the two
lines are the strongest lines in the spectrum of collision
dominated low-density plasma. For consistency, we describe the
property of the ratio $R_1$ briefly, and the model considered in
our calculation. The ratio $R_1$ is defined as
\begin{eqnarray}
R_1 & = & \frac{I(\lambda50.524~{\rm \AA\,})}{I(\lambda50.691~{\rm
\AA\,})+I(\lambda50.703~{\rm \AA\,})}
\end{eqnarray}

Radiative rates, and electron impact excitation as well as
de-excitation among 320 levels of Si X have been included in the
calculation of line intensities at the steady state condition for
different electron densities. The energy level is replaced by
experimental value if it is available. Excitation rates by protons
of Foster, Keenan $\&$ Reid (1997) are also taken into account,
which are calculated by using the close-coupled impact parameter
method. In the observed spectra, the two lines (50.691~\AA\, and
50.703~\AA\,) are blended. So the sum of the intensities of these
two lines is used in definition and calculation.

Figure 3 shows the ratio as a function of the electron density
$n_e$ at three different logarithmic electron temperature
(log$T_e$(K)): 5.9, 6.1 and 6.3. The ratio is sensitive to the
electron density in the range from 10$^7$~cm$^{-3}$ to
10$^{10}$~cm$^{-3}$ which matches conditions that known to appear
in various stellar object. The effect of electron temperature on
the ratio appears to be important at a low-density limit
$n_e<5\times10^6$cm$^{-3}$, while the effect is negligible above
$2\times10^8$cm$^{-3}$ as shown in Fig. 3. This is an important
feature for density diagnostics. The line intensity at
50.703~\AA\, holds the sensitivity to the electron temperature and
density. At the high-density, it is negligible when compared with
that of line at 50.691~\AA\,, but it appears to be comparable at
low-density limit. So the slight sensitivity to the electron
temperature is present in the low-density region as shown in Fig.
3.

\section{Results and discussions}
\subsection{Diagnostic of the electron density} Based on the
measurements of line fluxes for individual lines of Si X, the
observed line ratios are obtained for our sample, which are listed
in Table 3. All these ratios are located in the density-sensitive
range ($R_1~\sim$6---2.7), so the electron density of X-ray
emitting layer of stellar coronae can be derived. For Capella, the
emission lines of Si~X appear to be weak, and one feature
(50.524~\AA\,) is strongly contaminated by emission line of Fe XVI
as shown in sect.2, so the observed ratio shows a relatively large
uncertainty.

By comparing observed ratios with the theoretical prediction,
electron densities of stellar coronae are derived for our sample
which are listed in Table 3. Figure 4 shows such comparison, in
which the solid line is prediction at the logarithmic electron
temperature 6.1 of maximum fraction of Si X in the ionization
equilibrium of Mazzatto et al. (1998), symbols with errors are
observed ratios of our sample. All the observed ratios are around
unity within uncertainty, which result in that the diagnosed
electron densities distribute in the range of
10$^8$---10$^9$~cm$^{-3}$. One component ($\alpha$ Cen B) of the
binary $\alpha$ Centauri shows a same ratio with that of Procyon,
while the other component shows a value close to that of
$\epsilon$~Eri. Therefore the electron densities for the two stars
share the same value as shown in Fig. 4. The work of Raassen et
al. (2003) also indicates the similar properties between $\alpha$
Cen B and Procyon. The ratio between {\it Ly}$\alpha$ of H-like O
VIII and resonance lines (or triplet) of He-like O VII for
$\alpha$ Cen B is very close to that of Procyon (see top pannel of
Fig. 9 in work of Raassen et al. (2003)).

\subsection{Discussions}
A systematic analyses of extreme ultraviolet spectra for 28
stellar coronae performed by Sanz-Forcada et al. (2003), reveals
that the stellar coronae consistent with two basic classes of
magnetic loops: solar-like loops with maximum temperature around
log$T_e$(K)$\sim$6.3 and lower electron densities
($n_e\geq10^9$--10$^{10.5}$cm$^{-3}$), and hotter loops peaking
around log$T_e$(K)$\sim$6.9 with higher electron densities
($n_e\geq10^{12}$cm$^{-3}$). For the hotter temperature region,
the electron densities are usually derived from highly charged
iron ions with higher peak temperature in the ionization
equilibrium~(Mazzotta et al. 1998). Sanz-Forcada et al. (2003)
derived the electron densities are more than $10^{12}$cm$^{-3}$
from EUV spectra observed with EUVE satellite. An upper limit
($5\times10^{12}$cm$^{-3}$) for stellar coronae was concluded by
Ness et al. (2004) from carbon-like iron lines with fluxes
extracted from Low Energy Transmission Grating Spectrometer
(LETGS) spectra. From resolved triplets of He-like magnesium and
silicon, upper limits $7\times10^{11}$ and
$1\times10^{12}$cm$^{-3}$ for Capella were derived by Canizares et
al. (2000) from the HEG spectra. Through de-blending the
contamination for Ne~IX triplet, Ness et al. (2003) place an upper
limit of 2.0$\times10^{10}$cm$^{-3}$ for Capella.

Ness et al. (2002, 2004) further made a detailed investigation for
the coronal density using He-like triplet, and found that the
derived densities from He-like C, N and O do not deviate from
low-density limits for inactive stars. For the cool X-ray emitting
region, O~VII triplet is extensively used to derive the electron
density for stellar coronae because of its high SNR around
21~\AA\,. Whereas the C V triplet forms at a much cooler emitting
layer which should carry out information in this emitting layer.
For inactive and some active stars, the C V triplet can be
detected and the corresponding electron density has been published
in literatures(Ness et al. 2001). In ionization equilibrium of
Mazzotta et al. (1998) condition, the temperature (1.26~MK) of
maximum fractional abundance of Si X is close to that derived from
H- and He-like C. Therefore the electron densities from the two
approaches should be compatible with each other. Here we make a
comparison between our results and the values constrained by Ness
et al. (2001; 2002) for our sample, as shown in Fig. 5.

The figure reveals that our results are consistent with those of
Ness et al. (2001, 2002) within 1$\sigma$ statistical uncertainty,
which suggests that electron densities derived from spectra of
ions with similar formation temperature are compatible with each
other. For three inactive stars, the density derived from the two
methods shows an excellent agreement. The uncertainties of our
results are much smaller than those constrained by He-like C V.
The large uncertainty in the electron densities derived from
He-like C V is due to a faint inter-combination line presented,
which results in that the observed ratio $i/f$ between the
inter-combination~($i$) and forbidden~($f$) lines located near the
low-density limit. The flux measurement for $i$ line is strongly
depended on the determination of continuum level, so a relatively
larger uncertainty appears in constrained electron densities. For
selected lines in this study, no faint emission lines are
involved. The line fluxes of the two features are comparable for
these inactive stars. The observed ratios are close to unity,
which corresponds to the density-sensitive region. In case of
$\alpha$ Cen A (G2V), only upper limit of density can be obtained
from C V triplet because a faint inter-combination line is
present. However present work further constrains the mean electron
density for the cooler layer of the star.

In case of Capella, the derived electron density also agrees with
that derived from C V, whereas our result shows a larger error
bar. One reason is due to that the emission line of Si X at
50.524~\AA\, is severely contaminated by line of Fe XVI at
50.555~\AA\,. Another possible reason may be the determination of
continuum level, because the emission lines of Si~X have become
weak for active stars. For $\epsilon$~Eri, the electron density is
not available from C V because the forbidden line is severely
contaminated. The electron density of 1.29$\times10^9$~cm$^{-3}$
constrained by Si X could be used to represent the density in C V
forming region.

\section{Conclusions}
For inactive stars, such as Procyon and $\alpha$ Cen A and B,
emission lines of Si X at 50.524~\AA\, and 50.691~\AA\, are the
prominent lines with high SNR in the wavelength range
35---70~\AA\,. Our previous paper indicates that the ratio $R_1$
between the two lines is sensitive to electron density, while it
is insensitive to electron temperature. This is a good feature for
diagnostic of the electron density for hot astrophysical and
laboratory plasmas. In the low-density limit region, the obvious
temperature effect of the ratio is due to contamination by
$3d\to2p$ transition of Si X at wavelength 50.703~\AA\,. Compared
with the intensity of line at 50.691~\AA\,, the line intensity at
50.703~\AA\, steeply decreases with increasing the electron
density. In case of Capella, the line at 50.691~\AA\, is clearly
detected, while the line at 50.524~\AA\, is severely contaminated
by line of Fe XVI. Based on the conclusion of Argiroffi et al.
(2003) that the emission of Capella is compatible with a constant
source, we use the EMD derived by Audard et al. (2001) and line
emissivity included in APEC1.3.1 model to estimate the
contribution of Fe XVI line at 50.555~\AA\,. About 62\% flux is
estimated to be from the contribution of Fe XVI, the remaining
flux is from Si X line at 50.524~\AA\,. The contribution of Fe XVI
line at 50.367 is simultaneously estimated, which contributes
about 73\%. When the EMD is folded with emissivities of Fe XVII
lines, the estimated flux of Fe XVII lines around 50.367~\AA\, is
0.68$\times10^{-4}$photon~cm$^{-2}$~s$^{-1}$ which is consistent
with the remaining flux within statistical error. Applying the
procedure to $\epsilon$~Eri, we conclude that the contributions of
Fe XVI lines at 50.36~\AA\, and 50.54~\AA\, are about 53\% and
39\%, respectively. So the flux of Si X line at 50.524~\AA\, is
about 0.64$\times10^{-4}$~photon~cm$^{-2}$~s$^{-1}$.

The observed ratios of our sample are close to the unity which
corresponds the density-sensitive range of $R_1$. Comparison of
our results with those derived from He-like C V shows a good
agreement. Moreover, the derived electron densities and error bars
are similar for the inactive stars, but the error bars are large
for the active stars. In conclusion, the emission lines of Si X
have potential density-diagnostic application in stellar coronae
and other X-ray sources.

\begin{acknowledgements}
This work was supported by the National Natural Science Foundation
under Grant No. 10433010 and No. 10403007, as well as the Chinese
Academy of Sciences under Grant No. KJCX2-W2.
\end{acknowledgements}

\begin{table*}
\centering
    \caption[I]{Summary of stellar properties and measurement of X-ray
    luminosity (in range of 5---175~\AA\,) for the stars. t$_{obs}$ denotes the exposure time of observations.}
    \vspace{0.2cm}
      \[
      \begin{array}{lccccccccc} \hline\hline
{\rm star}  & {\rm HD} & {\rm ObsID} & {\rm t_{obs}} & {\rm
Spectr. Type} & {\rm distance^a} & T{\rm ^a_{eff}} & {\rm log({\it
L}_{bol})} & {\rm R_{\star}^a} & L{\rm _X^b} \\
 & & & {\rm ks} & & {\rm pc} & {\rm K} & {\rm erg/s} & {\rm [{\it R}_{\sun}]} &
 10^{28}{\rm erg/s} \\
\hline {\rm Procyon} & 61421 & 63 & 69.6 & {\rm F5.01V-V} & 3.5 &
6540 & 34.46 & 2.06 & 2.43\\
{\rm \alpha~Cen~A} & 128620 & 29 & 79.5 & {\rm G2.0V} & 1.34 &
5780 & 33.77 & 1.23 & 0.65 \\
{\rm \alpha~Cen~B} & 128621 & 29 & 79.5 & {\rm K0.0V} & 1.34 &
5780 & 33.28 & 0.8 & 0.52 \\
{\rm Capella} & 34029 & 1248 & 84.7 & {\rm G1.0III/K0.0I} & 12.94
& 5850 & 35.71 & 9.2/13 & 255 \\
{\rm \epsilon~Eri} & 22049 & 1869 & 105.3 & {\rm K2.0V} & 3.22 &
4780 & 33.11 & 0.81 & 20.9 \\ \hline
         \end{array}
      \]
\flushleft{$^a$From Ness et al. (2004) \hspace{1cm} $^b$From Ness
et al. (2002).}
      \end{table*}

\begin{table*}
\centering
    \caption[I]{Line fluxes (in unit of 10$^{-4}$photon~cm$^{-2}$~s$^{-1}$)
    of prominent lines in wavelength range 50---55~\AA\,
    for stars: Procyon, $\alpha$ Cen A$\&$B, Capella and $\epsilon$~Eri. The values
    in parentheses are the errors (in percent) with 1$\sigma$ uncertainty. }
    \vspace{0.2cm}
      \[
      \begin{array}{lcccccc} \hline\hline
  {\rm Ions}&{\rm \lambda_{theo}~(\AA)}  & {\rm Procyon} & {\rm
      \alpha~Cen~A} & {\rm \alpha~Cen~B} & {\rm Capella} & {\rm \epsilon~Eri} \\ \hline
{\rm Fe~XVI}& 50.350 & - & - & - & 4.01(37)^a & 1.31(15)^a\\
{\rm Si~X}  & 50.524 & 1.68(15) & 0.99(10) & 0.89(12) & 0.85(23)^a & 0.64(14)^a \\
{\rm Fe~XVI}& 50.555 & - & - & - & 2.07(28)^a & 0.40(14)^a\\
{\rm Si~X}  & 50.691 & 1.30(14) & 0.91(11) & 0.78(10) & 1.14(18) & 0.77(13) \\
{\rm Fe~XVI}& 54.142 & - & - & - & 2.57(40) & 0.76(14)\\
{\rm Fe~XVI}& 54.728 & - & - & - & 5.67(54) & 1.33(16)\\
\hline
         \end{array}
      \]
\flushleft{$^a$  blended lines}
      \end{table*}

\begin{table*}
\centering
    \caption[I]{Observed line intensity ratios for our sample, and diagnosed electron densities (in log$n_e$/cm$^{-3}$), together with
     the statistic errors within 1$\sigma$. }
      \[
      \begin{array}{lccccc} \hline\hline
   & {\rm Procyon} & {\rm \alpha~Cen~A~(G2V)} & {\rm \alpha~Cen~B~(K1V)} & {\rm Capella} & {\rm \epsilon~Eri}\\ \hline
{\rm Ratio} & 1.29\pm0.25 & 1.09\pm0.24 & 1.31\pm0.39 & 0.73\pm0.48 & 0.83\pm0.32 \\
{\rm log}n_e~{\rm (cm^{-3})} & 8.61^{+0.24}_{-0.20} &
8.81^{+0.27}_{-0.23} & 8.60^{+0.39}_{-0.32} & 9.30_{-0.48} &
9.11^{+1.40}_{-0.38} \\ \hline
         \end{array}
      \]
      \end{table*}

\begin{figure*}
\centering
\includegraphics[width=16.5cm,height=14cm]{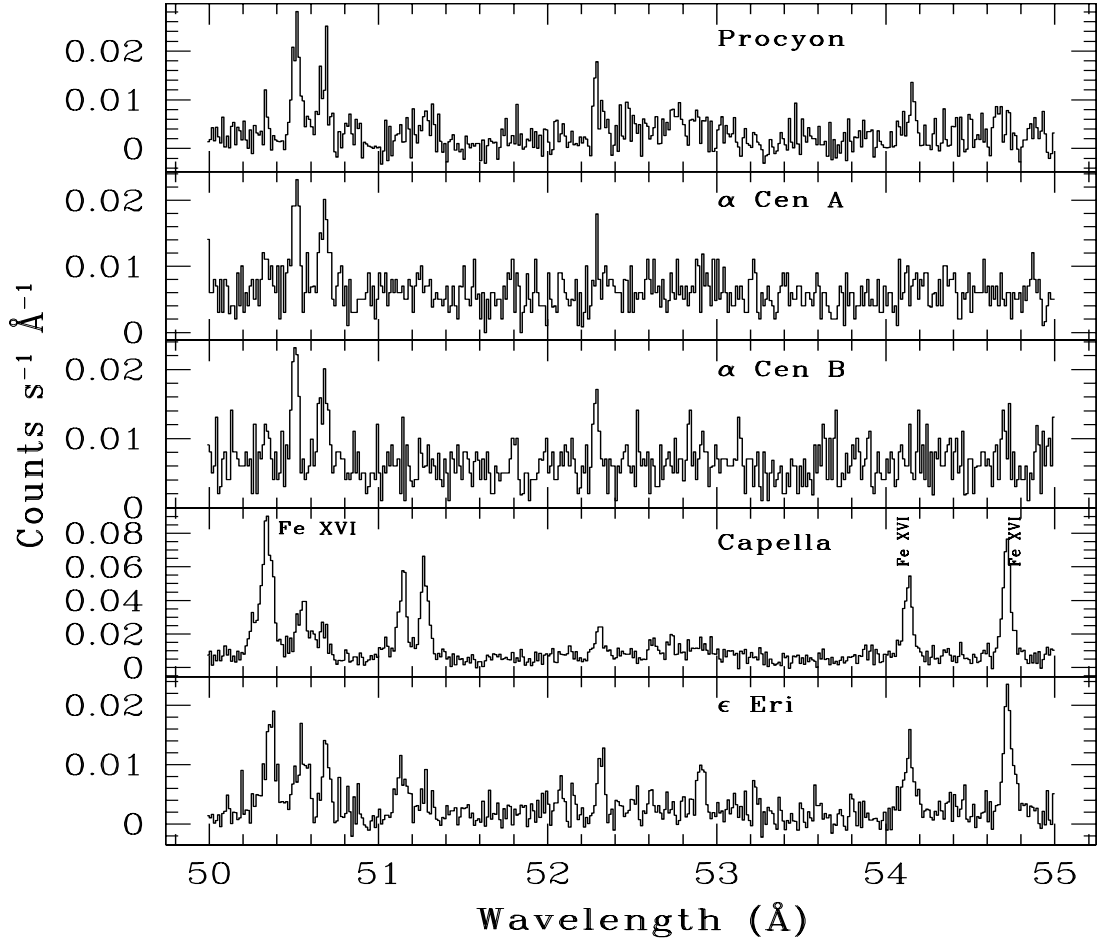}
\caption{Extracted LETGS spectra of our sample in wavelength range
50---55~\AA\,. HRC-S instrument was used for all observations. In
the forth panel (from top), prominent lines from Fe XVI are
labelled. }
\end{figure*}

\begin{figure}
\centering
\includegraphics[width=9cm,height=8.5cm]{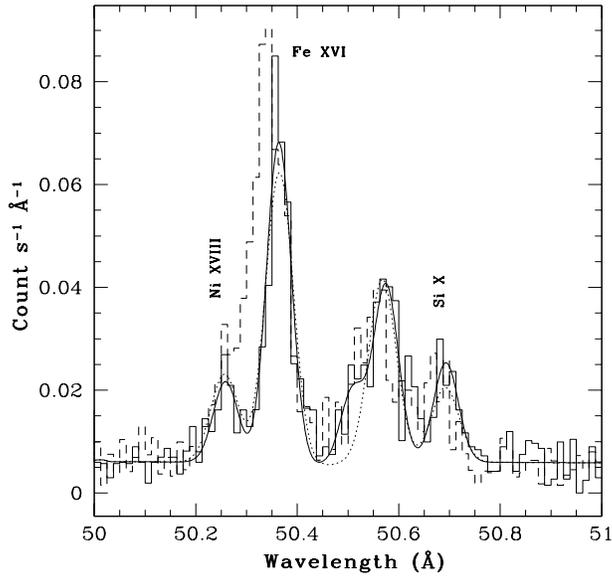}
\caption{Spectra of Capella observed with ACIS-S (solid histogram)
and HRC (dashed histogram) instruments in wavelength range
50---51~\AA\, and best-fit spectra of ACIS-S spectrum with five
(solid smooth line) and four (dotted line) Gaussian profiles,
together with a constant representing background and
(pseudo)continuum. }
\end{figure}

\begin{figure}
\centering
\includegraphics[width=6.5cm,angle=-90]{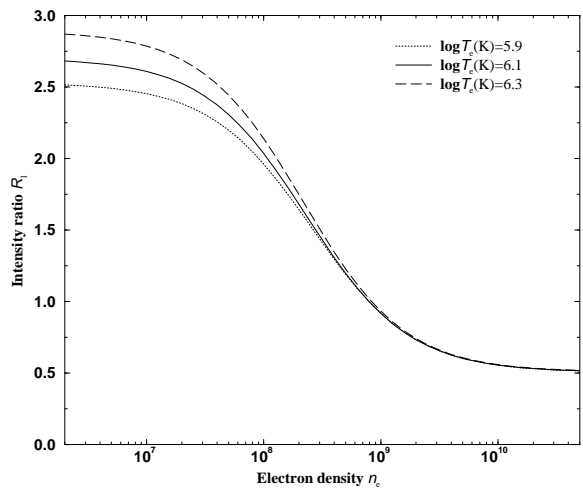}
\caption{Line intensity ratio $R_1$ as a function of the electron
density $n_e$ at three different electron temperature
(log$T_e$(K)): 5.9, 6.1 and 6.3.}
\end{figure}

\begin{figure}
\centering
\includegraphics[width=6.5cm,angle=-90]{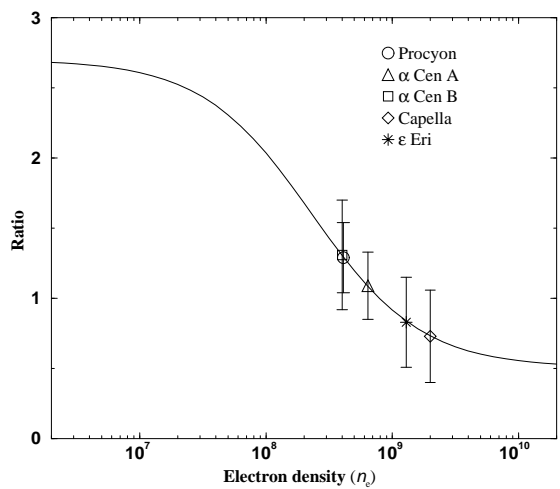}
\caption{Observed line intensity ratio (symbols with 1$\sigma$
error bars) for our sample and theoretical prediction at the
logarithmic electron temperature log$T_e$(K)=6.1 (solid line).}
\end{figure}

\begin{figure}
\centering
\includegraphics[width=6cm,angle=-90]{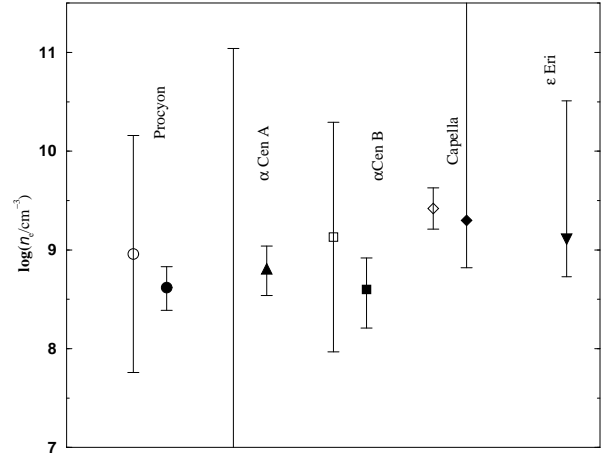}
\caption{The electron density derived from ratio $R_1$ of Si X
(filled symbols with errors) and from C V triplet (open symbols
with errors).}
\end{figure}

\end{document}